\documentclass[journal=jacsat,manuscript=preprint,layout=onecolumn]{achemso}
\usepackage{amsmath, amssymb, amsfonts, bm, textcomp}
\usepackage{float, placeins, tikz, graphicx}
\usepackage{multirow, array}

\newcommand{\bDGrs}{\ensuremath{\beta\Delta G_{\rm sol}\left(r\right)}\,\,}

\newcommand{\DGrs}{\ensuremath{\Delta G_{\rm sol}\left(r\right)}\,\,}
\newcommand{\DGrsp}{\ensuremath{\Delta G_{\rm sol}\left(r\right)}}

\newcommand{\DGs}{\ensuremath{\bar{\Delta G_{\rm sol}}}\,\,}
\newcommand{\DGsp}{\ensuremath{\bar{\Delta G_{\rm sol}}}}
\newcommand{\DGsc}{\ensuremath{\bar{\Delta G_{\rm sol}}}}

\newcommand{\etal}{{\it et al.}}

\newcommand{\lcst}{\ensuremath{T_{\rm c}}}

\newcommand{\DGsa}{\ensuremath{{\Delta G_{{\rm sol},A}}}}
\newcommand{\DGsb}{\ensuremath{{\Delta G_{{\rm sol},B}}}}
\newcommand{\DGa}{\ensuremath{{\Delta G_{A}}}}
\newcommand{\DGb}{\ensuremath{{\Delta G_{B}}}}
\newcommand{\DGt}{\ensuremath{{\Delta G_{AB}}}}

\newcommand{\Rnp}{\ensuremath{{R_{\rm np}}}}

\newcommand{\Rg}{\ensuremath{{R_{\rm g}}}}
\newcommand{\scale}{\ensuremath{\mathcal{P}}}

\title{Theory of Solvation-Controlled Reactions in Stimuli-Responsive Nanoreactors}
\author{Stefano Angioletti-Uberti} 
\affiliation{Institut f{\"u}r Physik, Humboldt-Universit\"{a}t zu Berlin, Newtonstr.~15, Berlin, Germany}
\affiliation{ Helmholtz-Zentrum~Berlin, Hahn-Meitner-Platz~1, Berlin, Germany}
\email{sangiole@physik.hu-berlin.de}
\author{Yan Lu} 
\affiliation{ Helmholtz-Zentrum~Berlin, Hahn-Meitner-Platz~1, Berlin, Germany}
\author{Matthias Ballauff} 
\affiliation{Institut f{\"u}r Physik, Humboldt-Universit\"{a}t zu Berlin, Newtonstr.~15, Berlin, Germany}
\affiliation{ Helmholtz-Zentrum~Berlin, Hahn-Meitner-Platz~1, Berlin, Germany}
\author{Joachim Dzubiella}
\affiliation{Institut f{\"u}r Physik, Humboldt-Universit\"{a}t zu Berlin, Newtonstr.~15, Berlin, Germany}
\affiliation{ Helmholtz-Zentrum~Berlin, Hahn-Meitner-Platz~1, Berlin, Germany}
\email{Joachim.dzubiella@helmholtz-berlin.de}

\begin{document}

\begin{abstract}

Metallic nanoparticles embedded in stimuli-responsive polymers can be regarded as nanoreactors since their catalytic activity 
can be changed within wide limits: the physicochemical properties of the polymer network can be tuned and 
switched by external parameters, e.g. temperature or pH, and thus allows a selective control of reactant mobility and concentration close
to the reaction site.  
Based on a combination of Debye's model of diffusion through an energy landscape and a two-state model for the polymer, here we develop an analytical expression for the observed reaction rate constant $k_{\rm obs}$. 
Our formula shows an exponential dependence of this rate on the solvation free enthalpy change \DGs, a quantity which describes the partitioning of the reactant in the network versus bulk. Thus, changes in \DGsc, and not in the diffusion coefficient, will be the decisive factor affecting the reaction rate in most cases. A comparison with recent experimental data on switchable, thermosensitive nanoreactors demonstrates the general validity of the concept.

\end{abstract}

\maketitle

\section{INTRODUCTION}

Metallic and oxidic nanoparticles have been the subject of intense studies in recent decades because of their catalytic activity.\cite{Astruc2008,Hutchings2012} For example, gold becomes an active catalyst for oxidation reactions when divided down to the nanoscale.\cite{Haruta2003,Hutchings2005,Thompson2007} Titania nanoparticles are also highly active catalysts\cite{Yurdakal2014,Graciani2014} and there is a large number of other nanometric systems with promising catalytic properties.\cite{Astruc2008} Use in catalytic reaction requires a simple and secure handling of nanoparticles by a suitable macromolecular carrier system, in particular when working in solution. Such a carrier system should not impede the catalytic activity of the nanoparticles but keep them firmly stabilized throughout the reaction and the subsequent work-up. Reactants should also be able to quickly enter and leave the carrier. A catalytic system  thus defined acts in many ways as a nanoreactor, that is, it contains and shelters the catalytic reaction. Examples of these nanoreactors are given, e.g., by dendrimeric systems\cite{Crooks2001} or those based on spherical polyelectrolyte brushes.\cite{Ballauff2007}

In recent years, a new class of carrier systems is emerging that can be termed active nanoreactors.\cite{Lu2006c,Lu2011,marzan-system,marzan1,lu-nanoreactors2,angewandte,Horecha2014,Chen2014} Here the nanoparticles are embedded in a polymer gel that reacts to external stimuli. In solution, diffusion within the gel may be manipulated by parameters such as temperature or pH. The best-studied examples of such active carriers are colloidal gels made from crosslinked poly(N-isopropylacrylamide) (PNIPAM) that undergo a volume phase transition at 32$^o$C. It has been demonstrated that the catalytic activity of metal nanoparticles embedded in such gels can be manipulated using temperature as the external stimulus.\cite{Lu2006c} 

In particular, a simple model system has been prepared by enclosing a single gold nanoparticles in a hollow PNIPAM sphere, a so-called yolk-shell architecture.\cite{angewandte} This active carrier in aqueous phase provides the ideal means to investigate the manipulation of the activity of a nanoparticle in a stimuli-responsive nanoreactor. It has been demonstrated that the reactivity of the nanoparticle can be switched depending on the hydrophilicity of the reactants.\cite{angewandte} These findings open a new way to introduce selectivity into the catalysis with nanoparticles. Another model system, introduced by Liz-Marz{\'a}n and co-workers \cite{marzan-system}, is made up by a single gold particle embedded in the middle of a colloidal PNIPAM sphere, a core-shell architecture. For this system, Carregal-Romero {\it et al.} demonstrated that the reduction of hexacyanoferrate (III) ions by sodium borohydride is slowed down drastically above the temperature of the volume transition, that is, when the gel has shrunken and most of the water has been expelled.\cite{marzan1} These workers explained this decrease of the rate constant by the decrease in the reactant's diffusion coefficient in the gel in its dense, collapsed state using a two-state model for the gel. In this way Carregal-Romero {\it et al.}  presented the first theory of the kinetics of catalysis by nanoparticles in such an active nanoreactor.\cite{marzan1}

Studying the reactivity of nanoparticles in the condensed phase requires a model reaction that allows us to obtain kinetic parameters related to catalysis with the greatest possible precision. Based on the pioneering work of Pal {\it et al.}\cite{Pradhan2002} and Esumi {\it et al.}, \cite{Esumi2002}  we\cite{Mei2005,Schrinner2009,Wunder2010,Wunder2011,Kaiser2012,Gu2014} and many others\cite{Xia2010,Zeng2010,Antonels2013,Nemanashi2013,Zhang2014A,Zhang2014B,Chi2014,Li2014,Chen2014,Shah2014, Herves2012} have demonstrated that the reduction of nitroarenes and especially of nitrophenol to 4-aminophenol by borohydride ions in aqueous phase fulfills all the requirements for such a model reaction.\cite{Herves2012} 

Using this model reaction, Wu {\it et al.} were able to show that the reactivity of the rather hydrophobic nitrobenzol is even increased when raising the temperature above the temperature of the volume transition.\cite{angewandte} This finding cannot be rationalized anymore in terms of a changed diffusion coefficient for the reactants. Wu {\it et al.} called attention on the fact that the thermodynamic interaction of the reactants with the PNIPAM network must be considered in this case, and used the Debye-Smoluchowski theory of diffusion on an energy landscape,\cite{debye42,smoluchowsky} to explain the observed results.

Here we present the full theory of nanoreactors that combines our previous analysis\cite{angewandte} with a two-state model\cite{zaccone, marzan1, heyda1, heyda2}  that takes into account the thermodynamic transition within the gel. The predictions of this model are compared to recent experimental data for both yolk-shell \cite{angewandte} and core-shell\cite{marzan1} systems. 
The remainder of the paper is organized as follows: First, in the Theory section we present a step-by-step derivation of our model, carefully stating its underlying assumptions. We then perform in the Discussion section a parameter study to highlight the effects of the various parameters in the model, and discuss the main predicted trends providing a comparison with available experimental data. Finally, a brief Conclusion will wrap up all results.

\section{THEORY \label{sec:theory}}

\subsection{Kinetics of surface reactions}

We start with the rate equation describing a
 chemical reaction of first order:
 \begin{equation}
\frac{ d  c_0 (t) }{dt} = - k_{\rm exp}  c_0 (t)
 \label{eq:k-exp}
 \end{equation}
 
 where $ c_0 $ is the concentration of reactants in solution, and $k_{\rm exp}$ is the experimental rate constant. 
By using a microscopic description we will arrive at an expression for the number of reactants transformed {\it by one} nanoreactor per unit time {\it 
at a constant bulk density $c_0$}, ending up with an equation of the form:
 
\begin{equation}
\frac{ dN }{dt} = k_{\rm obs}  c_0 .
\label{eq:k-ours}
\end{equation}
 
In order to connect $k_{\rm exp}$ and $k_{\rm obs}$, one needs to assume that nanoreactors do not interact with each other and can be treated independently (i.e., we can use a cell model, as explained in more details elsewhere\cite{protein-kinetics}). Given that under relevant experimental settings low nanoreactors' concentrations are used, this condition should be typically satisfied. 
Within this approximation:

 \begin{eqnarray}
\frac{ dN }{dt} &=& k_{\rm obs}  c_0  \nonumber \\ 
\nonumber \\
\frac{1}{V_{\rm sol}} N_{\rm nano}\frac{ dN }{dt} &=&  \frac{1}{V_{\rm sol}} N_{\rm nano} k_{\rm obs}  c_0 \nonumber \\
\nonumber \\
\frac{d c_0 }{dt} &=&  \frac{1}{V_{\rm sol}} N_{\rm nano} k_{\rm obs}  c_0 
\end{eqnarray}
leading to
\begin{eqnarray}
k_{\rm exp} =  -\frac{1}{V_{\rm sol}} N_{\rm nano} k_{\rm obs} 
 \label{eq:k-exp2}
 \end{eqnarray}
 
 where $N_{\rm nano}$ is the total number of nanoreactors present in a volume $V_{\rm sol}$ of solution.
Eq.\,(\ref{eq:k-exp2}) provides a link from the rate $k_{\rm exp}$ typically used to describe experiments and the microscopic rate at which nanoreactors transform reactants.\\
Since the reaction catalyzed by the nanoparticle takes place directly on its surface, the reaction rate per total surface area of catalyst (i.e., total nanoparticle surface area) and per volume of solution \cite{Mei2005,Wunder2010,Wunder2011,Kaiser2012,Gu2014,Herves2012} is necessary to compare different systems in an unbiased manner, hence:
 
\begin{align}
 \tilde{k}_{\rm exp} &= \frac{k_{\rm exp}}{S_{\rm tot}/V_{\rm sol}}
 			=\frac{N_{\rm nano} k_{\rm obs}}{S_{\rm tot}}
			 =\frac{k_{\rm obs}}{S_0}
 \label{eq:k-exp3}
 \end{align}
 
 where $S_0$ is now the surface area of a single nanoparticle inside the nanoreactor. 
 Via Eq.\,(\ref{eq:k-exp3}), we see how this latter normalized rate constant $\tilde{k}_{\rm exp}$ is truly an intensive measure of the properties of a single-nanoreactor, 
 and is the quantity that should be used for comparison within different systems.
 We next proceed to describe our microscopic model for the calculations of $k_{\rm obs}$.

\subsection{Model}

\begin{figure}[htb]
\begin{center}
\includegraphics[width=0.50\textwidth]{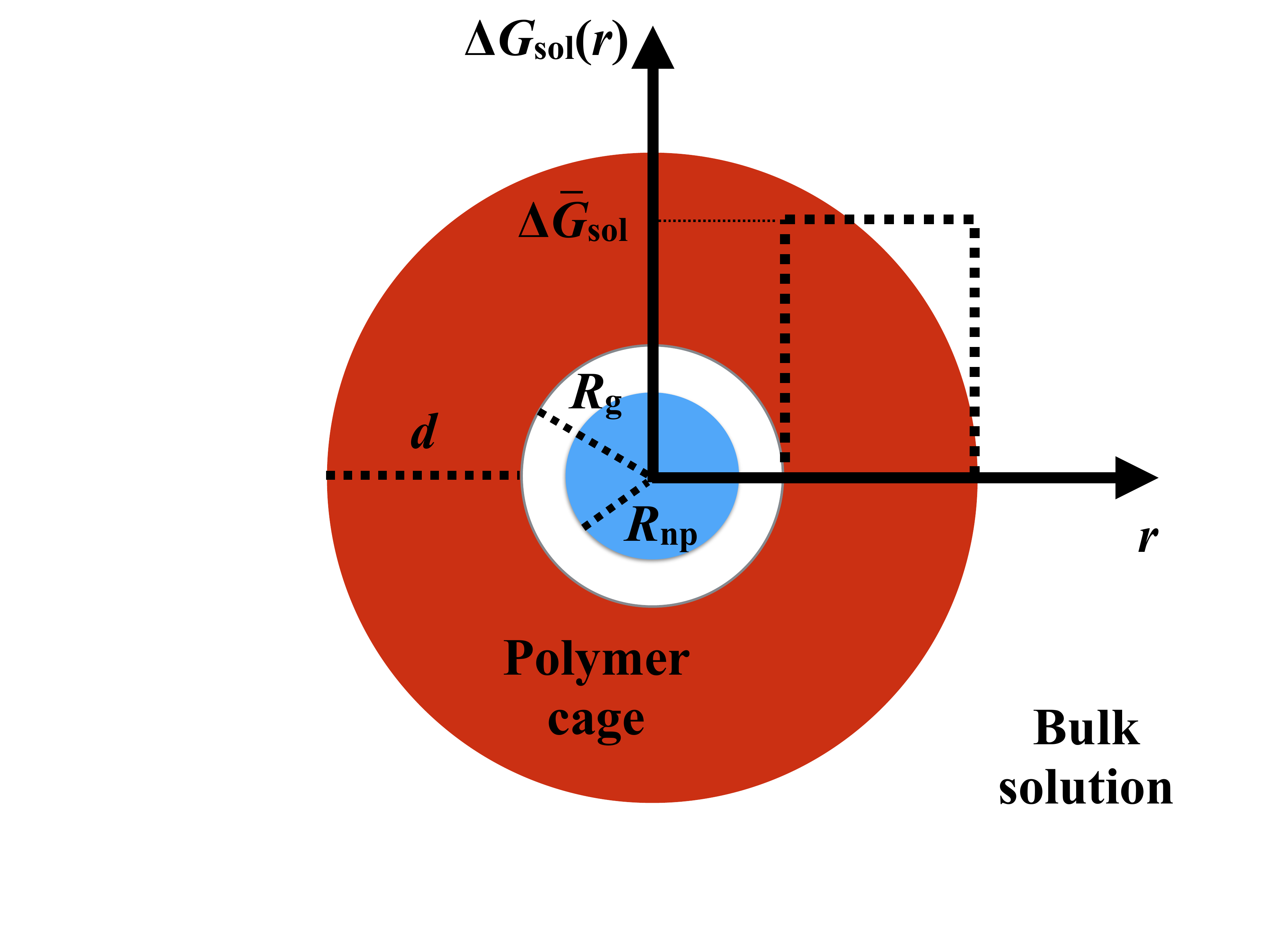}
\caption{\small{Schematic representation of a yolk-shell nanoreactor and central quantities for our theory. 
At the center core of the nanoreactor sits a metal nanoparticle (blue) of radius $R_{\rm np}$, embedded in a spherical 
polymer shell ('cage') of  inner radius $R_{\rm g}$ and outer radius $R_{\rm g}+d$. The reactants have to overcome a solvation free enthalpy barrier \DGrs to reach the nanoparticle depicted by the dotted lines. In our model, \DGrs is represented as a 
piecewise constant function, of value \DGs inside the polymer shell, and zero otherwise. In the case shown, 
\DGs is higher in the gel than in the bulk solution, representing ``polymer-phobic'' reactants, with an energetic penalty to absorb inside the shell, but the opposite situation can also occur.}}
\label{fig:system}
\end{center}
\end{figure}

A schematic representation of a yolk-shell nanoreactor is shown in Fig.\,\ref{fig:system}. For such a spherically symmetric system, we intend to model the rate for the catalytic reaction occurring at the surface of the nanoparticle, which requires taking into account the diffusion process that brings a reactant from the bulk solvent to the surface of the nanoparticle inside the nanoreactor's core.\cite{angewandte}  The kinetics of this process is governed by two key functions,  the diffusion coefficient of the reactants along the trajectory expressed by a diffusivity profile $D(r)$ and a thermodynamic free
enthalpy of solvation profile \DGrsp.  This local free enthalpy \DGrs measures the difference in the local excess chemical potential and the 'reservoir' chemical potential of the reactants far away from the nanoreactor.   Along its trajectory, the reactant encounters a different chemical environment in the gel, and its average interaction with it then results  in local $D(r)$ and \DGrs different than in bulk.

In order to keep the model as simple as possible, we take 
both $D(r)$ and \DGrs to be a step function centered in the polymer cage of the nanoreactor, of width equal to the shell width $d$ 
(see Fig.\,\ref{fig:system} for reference). This is a good approximation for spatially homogeneous gels that have been under consideration recently.\cite{angewandte} 
Thus, we get:
\begin{equation}
D(r) =
\begin{cases}
 D_0 ~~~{\rm for}~~ r<\Rg~~\mathrm{or}~~r>\Rg+d\\
 D_{\rm g} ~~~{\rm for}~~ \Rg \leq r \leq \Rg + d
\label{eq:DiffusionCoeff}
\end{cases}
\end{equation}
and 
\begin{equation}
\DGrs =
\begin{cases}
 ~~~0 ~~~~~~~{\rm for}~~ r<\Rg~~\mathrm{or}~~r>\Rg+d\\
 \DGs ~~~{\rm for}~~ \Rg \leq r \leq \Rg + d
\label{eq:DeltaG}
\end{cases}
\end{equation}
where \Rg\,\,is the inner radius of the polymer shell and $d$ its width, cf.~Fig.1. 
The choice of step functions basically implies that a reactant sees only two distinct environments, 
the interior of the polymer shell or the bulk solution.  Hence, in this form, \DGs can be identified with the difference in the 
solvation free enthalpy of the reactants in the gel with respect to that in the bulk solvent. As such, \DGs does not depend solely 
on the reactant-polymer interaction, but also on the solvent constituting the bulk solution. 
By measuring the equilibrium partitioning coefficient of the molecule in the gel $K_{\rm eq}$,
\DGs can be measured experimentally, or computed via atomistic simulations, with the following equation:
\begin{equation}
K_{\rm eq} =  c _{\rm g} /  c _{0} = \exp[-\beta\DGs]
\label{eq:partition}
\end{equation}
where $ c _{\rm g}$ and $ c _{0} $ are the equilibrium concentrations of reactant in the polymer gel and in the bulk solution, respectively, 
and $\beta=1/k_{\rm B} T$ is the thermal energy.\\
As for the free enthalpy, the two diffusion coefficients $D_{\rm g}$ and $D_0$ appearing in Eq.\,(\ref{eq:DiffusionCoeff}) can be identified as the bulk value in the polymer 
gel and solution, respectively.
The diffusion of solutes through hydrogels is a complex process\cite{amsden,masaro} and depends not only on the individual properties of the polymer, solvent, and solutes, 
such as size and concentration, but also on the particular interactions of the solute with the polymer network.  
In our case, we restrict the discussion to homogeneous gels with semiflexible polymers in the dilute and semi-dilute regimes with very small diffusing solutes
with a size comparable to a polymer monomer. 
In this case obstruction or hydrodynamic theories for small diffusing particles should better describe our problem.\cite{amsden,masaro} 
%
Regardless, the important point we want to emphasize is that all theoretical approaches to the diffusion process in our regimes predict a {\it decrease} of diffusion for
higher packing fractions. The detailed dependence of $D_{\rm g}$ on the system parameters is thus not of importance to discuss the qualitative features
of the temperature dependence of the reaction rate of nanoreactors.
%
%
\subsection{Coupling spatial diffusion and reaction: Debye-Smoluchowski approach \label{sec:debye}}

Once the diffusivity profile and the underlying free enthalpy on which the diffusing reactant must move to reach the surface of the nanoparticle are specified, it is 
possible to use an approach, initially developed by Smoluchowski \cite{smoluchowsky} and then extended by Debye, \cite{debye42} to calculate the diffusion-controlled part of the reaction rate in our system.  
This approach was first used to describe the rate of collision between charged ions in solution,\cite{debye42}  but the underlying physical picture is equivalent to our model.  
For the reader's convenience we provide a step-by-step derivation of the final equations in the Supplementary Information, and we also suggest to 
consult the reviews of Calef and Deutch \cite{deutch}  and Berg and von Hippel \cite{von-hippel} on diffusion-controlled reactions. The Debye-Smoluchowski model 
describes the rate at which a particle, driven by gradients in the chemical potential, diffuse from a bulk solution kept at  constant concentration $ c _0$ towards a fixed sink of radius $R_{\rm np}$. Close to the sink, 
a certain number per unit time of the molecules arriving are allowed to absorb ('react') as expressed by a surface reaction rate constant $k_{\rm R}$.  
Restated in our language, the sink is nothing but the nanoparticle inside the nanoreactor, and absorption into the sink means reacting in the proximity of the nanoparticle's surface. The total reaction rate $k_{\rm t}$, in units of reacting particles per unit time, arising from this model can be written as:

\begin{align}
\centering
k_{\rm t}^{-1} &=  k_{\rm R}^{-1} + k_{\rm D}^{-1}~~{\rm or} \nonumber \\
\tau_{\rm t} &= \tau_{\rm R} + \tau_{\rm D}
\label{eq:ktot-split-text}
\end{align}\\

where $k_{\rm R}$ and $k_{\rm D}$ are the so-called reaction and diffusion rates, respectively.  The times $\tau_t, \tau_{\rm R}\,\,{\rm and\,\,}\tau_{\rm D}$  defined in Eq.\,(\ref{eq:ktot-split-text}) 
are thus just the reciprocal of the corresponding rates, and should be interpreted as the effective time to diffuse to the sink $\tau_{\rm D}$, and to react once in the surface proximity in $\tau_{\rm R}$.  
Under these conditions the relation between the diffusivity and  free enthalpy landscapes $D(r)$ and $\Delta G_{\rm sol}(r)$, respectively,  and the diffusion rate $k_{\rm D}$ is given by the 
Debye-Smoluchowski expression (see the SI)
\begin{equation}
\centering
k_{\rm D} = 4 \pi  c _0 \left [ {\int_\Rnp^{\infty}\frac{\exp\left[\bDGrs\right]}{D(r) r^2} {\rm d}r}\right]^{-1}.
\label{eq:kdiff}
\end{equation}

\subsection{Solving Debye's equation for yolk-shell nanoreactors \label{sec:debye-applied}}

In $k_{\rm D}$, an integral of \DGrs over space appears, which can be analytically expressed using our simplified 
step-wise forms for $D(r)$ and \DGrs valid for nanoreactors, see Eq.\,(\ref{eq:DeltaG}), obtaining:

\begin{align}
k_{\rm D}^{-1} &= \tau_{\rm D} \\
		   &= \int_{R_{\rm np}}^{\infty} \frac{\exp[\beta \Delta G(r)]}{4 \pi r^2  c _{0}D(r)} = \nonumber \\
		   			&= \int_{R_{\rm np}}^{\Rg} \frac{1}{4 \pi r^2  c _{0} D_{0}} + \int_{\Rg}^{\Rg+d} \frac{\exp[\beta \Delta \bar G_{sol}]}{4 \pi r^2  c _{0}D_{\rm g}} + \nonumber \\
		   			&\quad \int_{\Rg+d}^{\infty} \frac{1}{4 \pi r^2  c _{0} D_{0}} \nonumber \\
		   			&= \tau_{\rm np} + \tau_{\rm g} + \tau_{\infty}
\label{eq:rate-split}
\end{align}\\

where again we split the typical time $\tau_{\rm D}$ into three different contributions: the effective time to arrive from the bulk solution to the gel, $\tau_{\infty}$, that to cross the gel  $\tau_{\rm g}$, and that to get to the surface of the nanoparticle once the gel has been crossed, $\tau_{\rm np}$.

Solving for the integrals in Eq.\,(\ref{eq:rate-split}) gives the following result:
\begin{align}
\tau_{\rm np} &= \frac{1}{ c _{0} 4 \pi D_{0}}\frac{\Rg-R_{\rm np}}{\Rg R_{\rm np}} \nonumber\\
\tau_{\rm g} &= \frac{\exp(\beta \Delta \bar G_{\rm sol})}{ c _{0} 4 \pi D_{\rm g}}\frac{d}{(\Rg+d)\Rg} \nonumber\\
\tau_{\infty} &= \frac{1}{ c _{0} 4 \pi D_{0} (\Rg+d)} 
\label{eq:rates-final}
\end{align}

For typical nanoreactors, $d \gg \Rg \approx R_{\rm np}$ ( which is exact when the reactor has a core-shell 
instead of a yolk-shell structure\cite{angewandte}). 
Considering also that $D_0>D_{\rm g}$, one can further simplify Eq.\,(\ref{eq:rates-final}) as:

\begin{center}
\begin{align}
\tau_{\rm np}& \approx 0 \\
\tau_{g}& \approx \frac{\exp(\beta \Delta \bar G_{\rm sol})}{ c _{0} 4 \pi D_{\rm g} \Rg} \gg \tau_{\infty},
\end{align}
\end{center}

Since the largest typical time, i.e. the slowest rate, dictates the final effective value for the reaction rate, 
the latter inequality implies:

\begin{center}
\begin{align}
k_{\rm D} \approx 1/ \tau_{\rm g} &=  4 \pi D_{\rm g}  c _{0} \Rg\exp\left(-\beta \Delta \bar G_{\rm sol}\right) \nonumber \\
					   & =  4 \pi  c _{0} \Rg \scale  
\label{eq:kdiff-nano}
\end{align}
\end{center}

where we introduced the quantity $\scale=D_{\rm g} \exp\left(-\beta \Delta \bar G_{\rm sol}\right) $, sometimes referred to as ``permeability'' 
(or inverse diffusive resistance),\cite{permeability1,Herves2012} which takes into account the 
compounded effect of the diffusion coefficient and the solvation free enthalpy.\\

The dependence of the surface reaction rate $k_{\rm R}$ on temperature can be simply modelled using an Arrhenius form to account for the 
effect of the activation energy necessary for the reaction to occur. 
In this case $k_{\rm R}$ (defined via solution of the Debye problem, see SI, Eq.\,(\ref{eq:kR-Debye}) is simply augmented by an exponential term, i.e:
\begin{equation}
k_{\rm R} \rightarrow k_{\rm R} \exp\left( -\frac{\Delta E_{\rm R}}{k_{\rm B}T}  \right),
\label{eq:surface-rate}
\end{equation}\\
where $\Delta E_{\rm R}$ is the activation energy of the surface reaction 
(see the discussion of this point in ref. \cite{Herves2012}).
%

\subsection{Coupling the Debye-Smoluchowski approach with a two-state-model \label{sec:2-states-model}}

The stimuli-responsive polymer gels used in active nanoreactors exhibit a volume transition between a swollen and a collapsed state at its lower critical solution temperature (LCST), \lcst.   
Since the transition can also be induced by changing pH or other external variables, for the sake of generality we describe the state of the gel as being a function of 
an unspecified general external parameter $\xi$, and represent it via a two-state model.\cite{zaccone, marzan1, heyda1, heyda2}  
More precisely, calling $\DGt(\xi) = \DGa(\xi) - \DGb(\xi)$ the difference of free enthalpy between the swollen ($A$) and collapsed ($B$) states, 
the Boltzmann-weighted probability to be in state $A$ is given by:
\begin{equation}
p_{\rm A}(\xi) = \frac{ \exp[-\beta \DGt(\xi) ] }{1 + \exp[-\beta \DGt(\xi) ]}.
\label{eq:probability}
\end{equation}
In a two-state model, the probability of the system being in the $B$ state is thus  $1-p_{\rm A}(\xi)$. 
If one now assumes that a molecule coming from the bulk solution towards the gel will see one of the two environments with a probability given by 
Eq.\,(\ref{eq:probability}), we arrive at the following equation for the total observed reaction rate:%
\begin{equation}
k_{\rm obs} = \langle k_{\rm t} \rangle =  p_{\rm A} k_{\rm t}^A + ( 1-p_{\rm A} ) k_{\rm t}^B.
\label{eq:k_obs}
\end{equation}
where $(k^{\alpha}_{\rm t})^{-1} = k^{-1}_{\rm R}(\Delta \bar{G}_{{\rm sol},\alpha})+ k^{-1}_{\rm D}(\Delta \bar{G}_{{\rm sol},\alpha},D_{{\rm g},\alpha})$, with $\alpha = A,B$. 
Eq.\,(\ref{eq:k_obs}) is equivalent to considering a situation where a reactant diffuses in a constant environment. 
Each environment will give rise to a different rate, and the observed one is just the thermodynamic average between the two.
This assumption, also implicitly made by Carregal-Romero \etal in their model,\cite{marzan1} becomes better the larger is the time it takes a system to swap 
between these two states of the gel compared with the average time a single reactant molecule takes to diffuse through 
the gel (which is of order $d^2/D_{\rm g}$, and should not be confused by $\tau_{\rm D}$).\\
\begin{figure}[H]
\begin{center}
\includegraphics[width=0.49\textwidth]{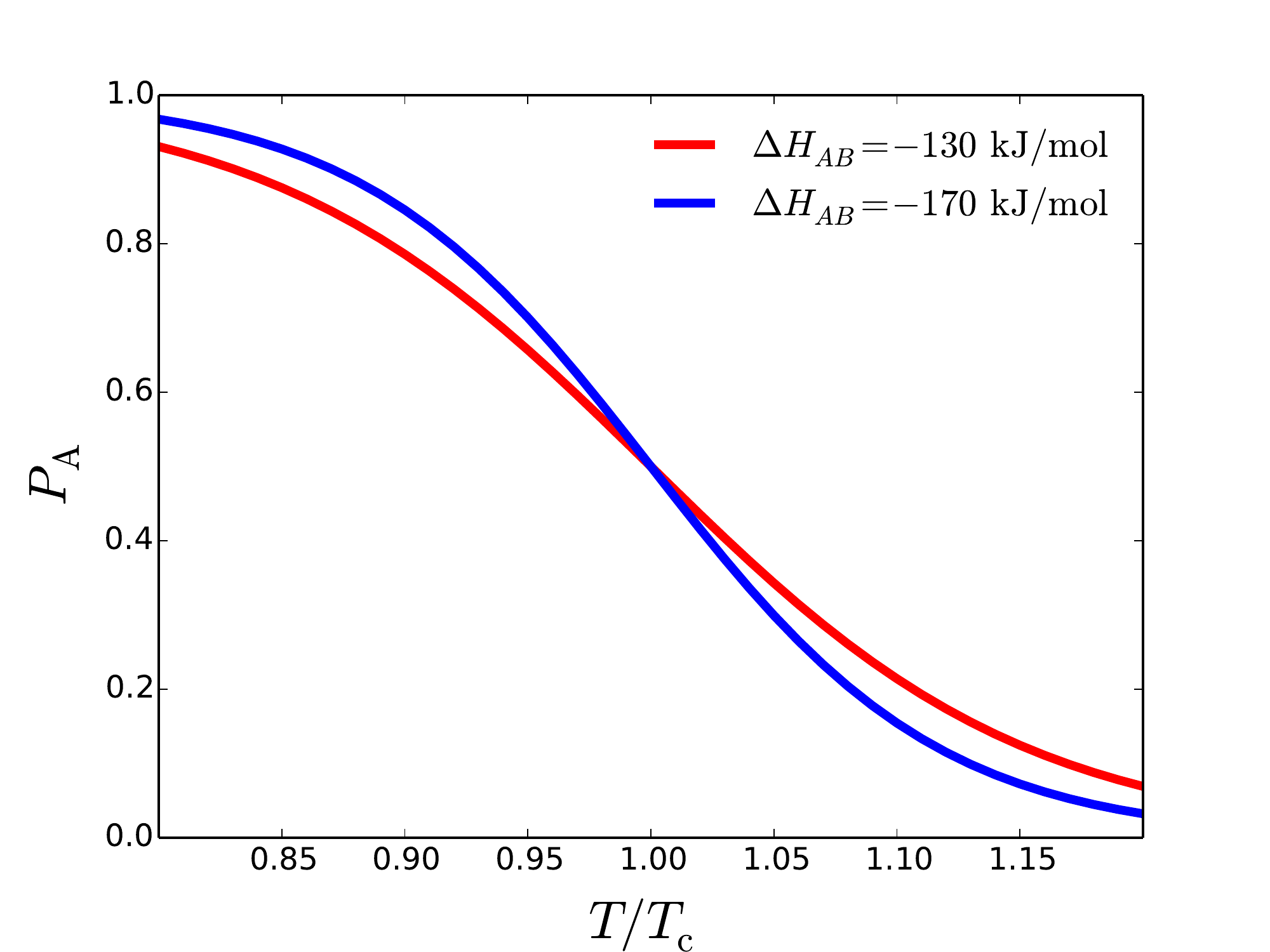}
\caption{Probability to find the system in the $A$ state, $P_{\rm A}(\xi)$ as a function of the control parameter $\xi$ in a thermodynamic two-states model, for two different
 values of the transition enthalpy $\Delta H_{AB}$. The latter is defined through its connection with the transition free enthalpy via the standard relation
$\DGt(\xi) = \Delta H_{AB} - T \Delta S_{AB}$. At the critical value for the control parameter $\xi$, $p_{\rm A}$ moves from 0 to 1, with increasing sharpness for increasing
$\Delta H_{AB}$\,\,(note that $p_{\rm B}$ is simply $1-p_{\rm A}$).}
\label{fig:transition}
\end{center}
\end{figure}

On the opposite limit instead, when the gel switches infinitely fast between the two states, 
a reactant crossing the gel would see an average environment, 
i.e., an average solvation free enthalpy and an average diffusion coefficient.\cite{hanggi} 
In this case, we have the following equation for the predicted rate:
\begin{align}
k_{\rm obs,fast} &= k_{\rm t}(\langle \DGs \rangle ,\langle D_{\rm g} \rangle),
\label{eq:k_obs-fast}
\end{align}
where, as for Eq.~\ref{eq:k_obs}, the two-state average $\langle .. \rangle$ of a quantity $X$ is again simply defined via the mean
\begin{align}
\langle X\rangle &= p_{\rm A}(T) X_{\rm A} + (1-p_{\rm A}(T))X_{\rm B}. 
\end{align}
\begin{figure}[h]
\begin{center}
\includegraphics[width=0.45\textwidth]{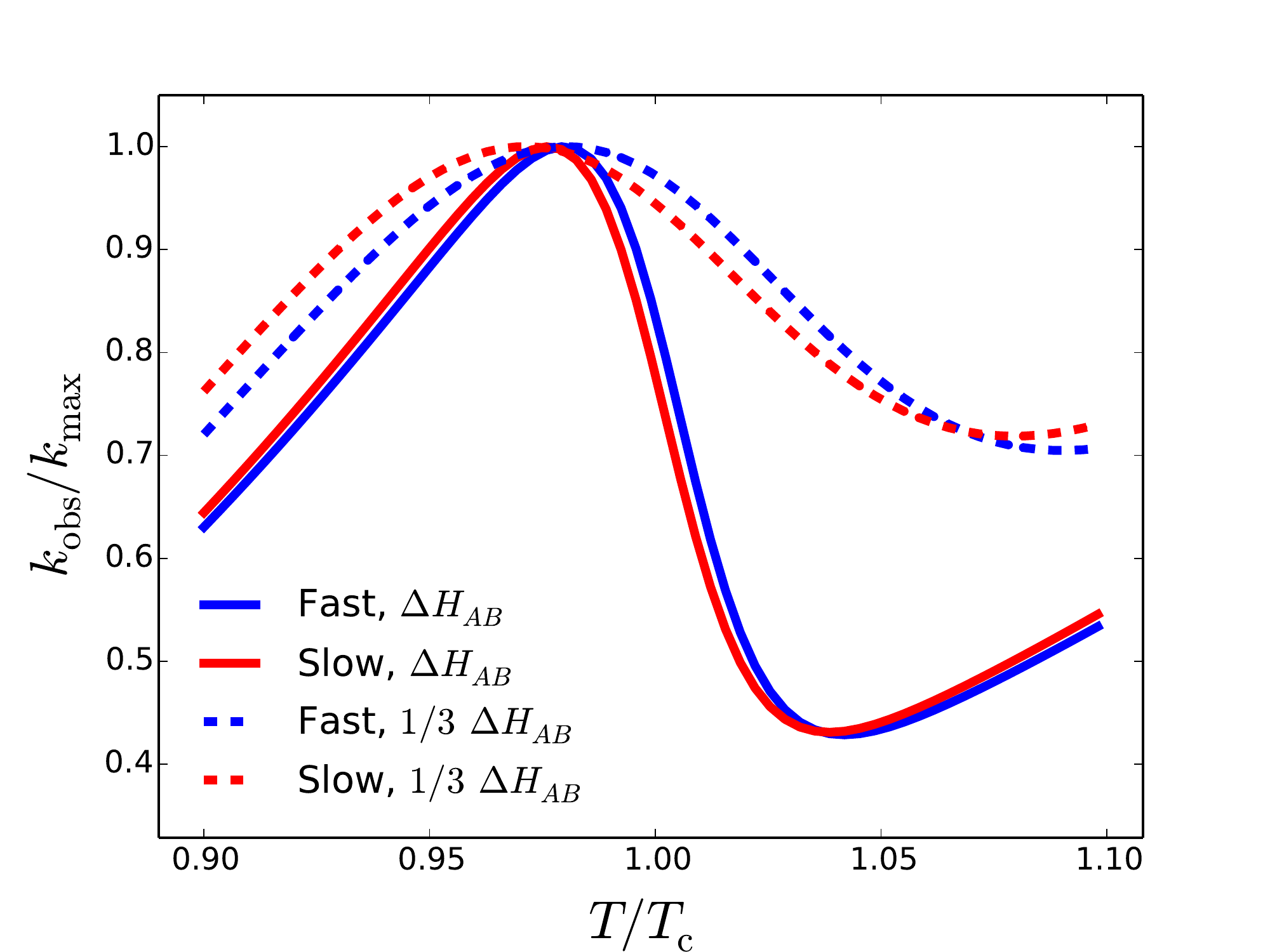}
\caption{Comparison of the reaction rate of a single nanoreactor calculated in the limit of infinitely fast (blue) and infinitely slow (red) swelling of the polymeric gel
i.e. using Eq.\,(\ref{eq:k_obs-fast}) and Eq.\,(\ref{eq:k_obs}), respectively, for two different values of the transition enthalpy $\Delta H_{\rm trans}$. The two limits describe qualitatively similar results, whereas quantitatively their agreement 
depends on the transition enthalpy, i.e. from the sharpness of the transition between the swollen and collapsed state as a function of the control parameter $\xi$ (here, $\xi$ is taken to be temperature). }
\label{fig:fast-vs-slow}
\end{center}
\end{figure}
There is a quantitative difference in the reaction rates calculated within these two different limits, as shown in Fig.\,\ref{fig:fast-vs-slow}, but both show 
the same qualitative trends.

We note, however, that in those cases where the swapping rate is finite (but not zero) and close to the crossing rate, 
resonance effects might give an additional signature close to the transition temperature (cf., a review on stochastic resonance\cite{resonance}). 
The investigation of those effects is out of scope of this paper and will be addressed in future studies. 

Let us now describe the reaction rate behavior resulting from Eq.\,(\ref{eq:k_obs}): close to the critical value for the parameter $\xi$, i.e., where $\DGt(\xi)=0$, 
the system continuously but sharply changes from the $A$ to the $B$ state (see Fig.\,\ref{fig:transition}).
These two states of the polymer cage represent different physicochemical environments, hence we expect the solvation free enthalpy \DGs for a reactant to also change markedly 
between them. Since \DGs enters exponentially in the observed reaction rate, Eq.\,(\ref{eq:kdiff-nano}), a sudden change upon polymer transition of \DGs 
will be reflected by a concurrent sudden jump in $k_{\rm obs}$. 
However, given that also the diffusion coefficient will jump from two distinct values, the overall sign of the change will depend on the  permeability $\scale$ and not simply \DGsp.  Despite this, we notice that this latter quantity enters \scale explicitly exponentially, whereas the diffusion coefficient only linearly, and it is thus expected to play a larger role. \\
Eq.\,(\ref{eq:k_obs}), together with the definitions in Eqs.~(\,\ref{eq:rate-split}) and (\ref{eq:rates-final}), allows us to clarify the connection between our 
model and the model presented by Carregal-Romero {\it et al}.\cite{marzan1} If we set $\DGs = 0$, that is, had we 
not accounted for the interaction between polymer and reactant,  Eq.\,(\ref{eq:k_obs}) would be exactly equivalent 
to their model, which describes the reaction rate purely in terms of diffusion.

Eq.\,(\ref{eq:kdiff-nano}) further suggests another insightful way to interpret our results. The diffusion rate $k_{\rm D}$ in both models can be written as $k_{\rm D} = 4 \pi R D  c (R_{\rm g})$, where $c (R_{\rm g})$ is the local concentration of the reactant in the gel. 
In the model by Carregal-Romero \etal,  $ c (R_{\rm g}) =  c _0$, i.e. the bulk concentration. In our model, since we consistently account for the 
thermodynamics of the system, we obtain $ c (R_{\rm g}) =  c _0 \exp[-\beta\DGs]$.

\begin{figure}[H]
\begin{center}

\begin{flushleft}
\leftskip = .2\hsize
A)\\
\end{flushleft}
\vspace{-1.cm}
\includegraphics[width=0.45\textwidth]{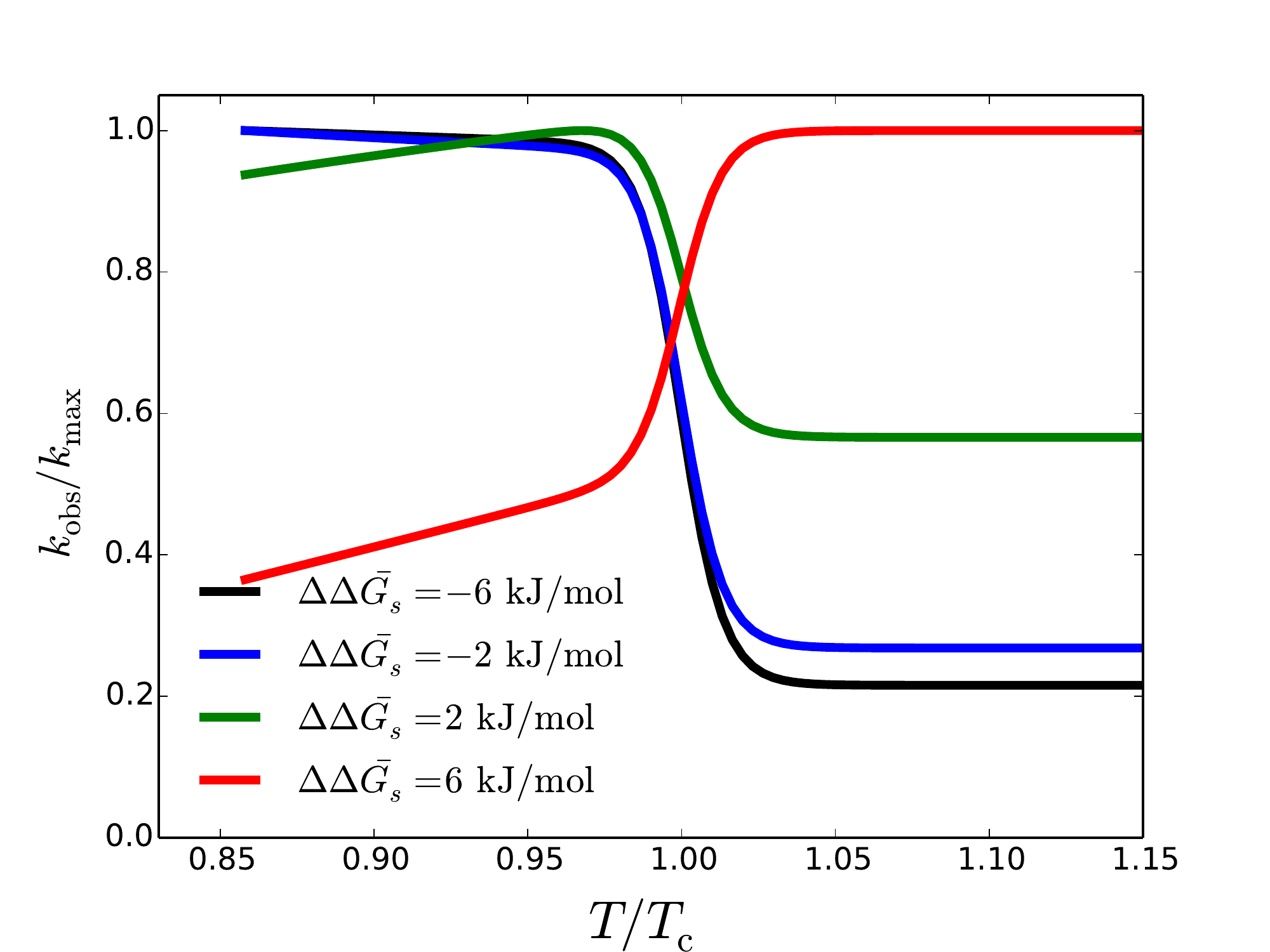}\\
\vspace{0.35cm}

\begin{flushleft}
\leftskip = .2\hsize
B)\\
\end{flushleft}
\vspace{-1.5cm}
\includegraphics[width=0.45\textwidth]{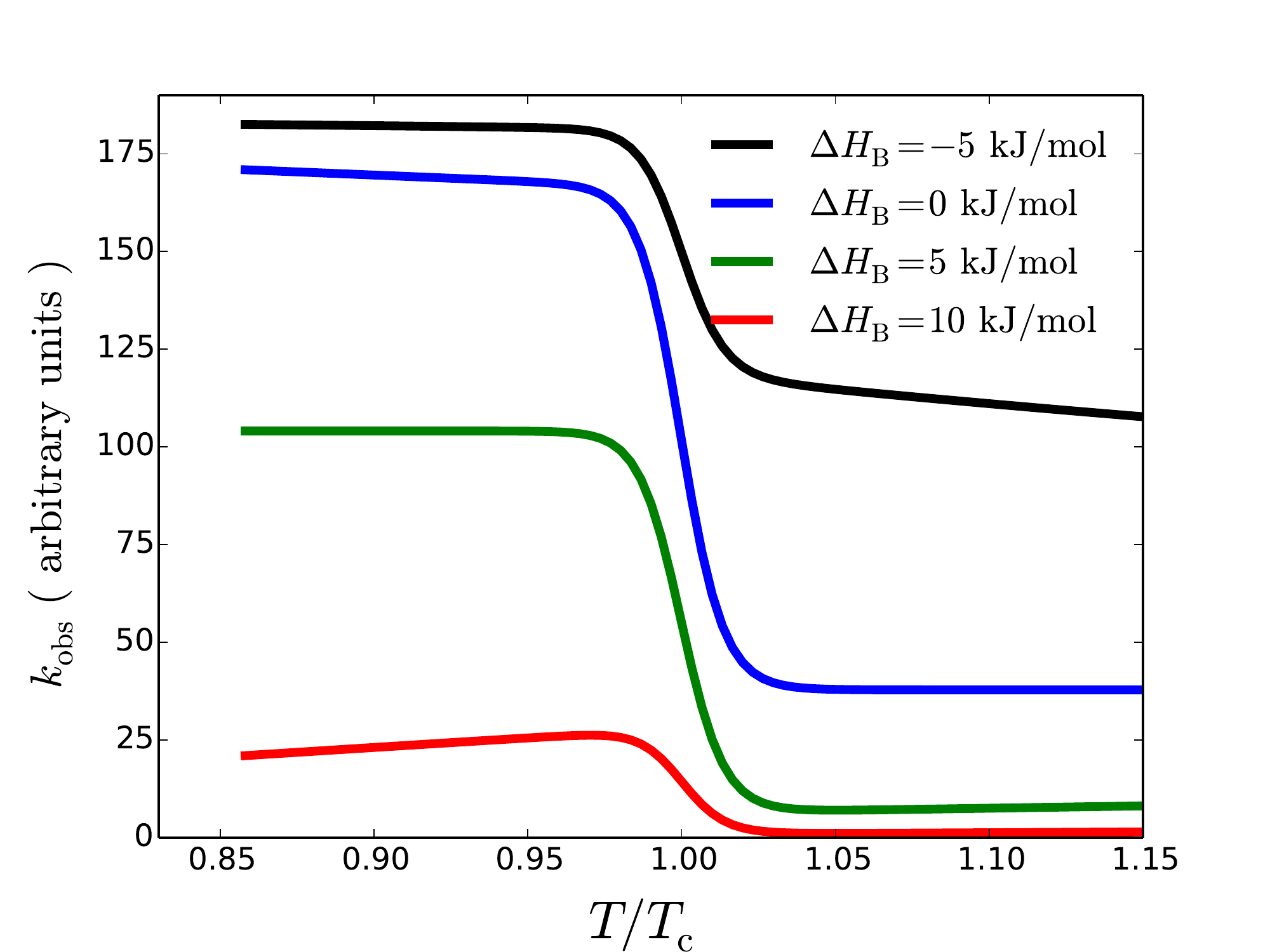}\\
\vspace{0.35cm}

\begin{flushleft}
\leftskip = .2\hsize
C)\\
\end{flushleft}
\vspace{-1.5cm}
\includegraphics[width=0.45\textwidth]{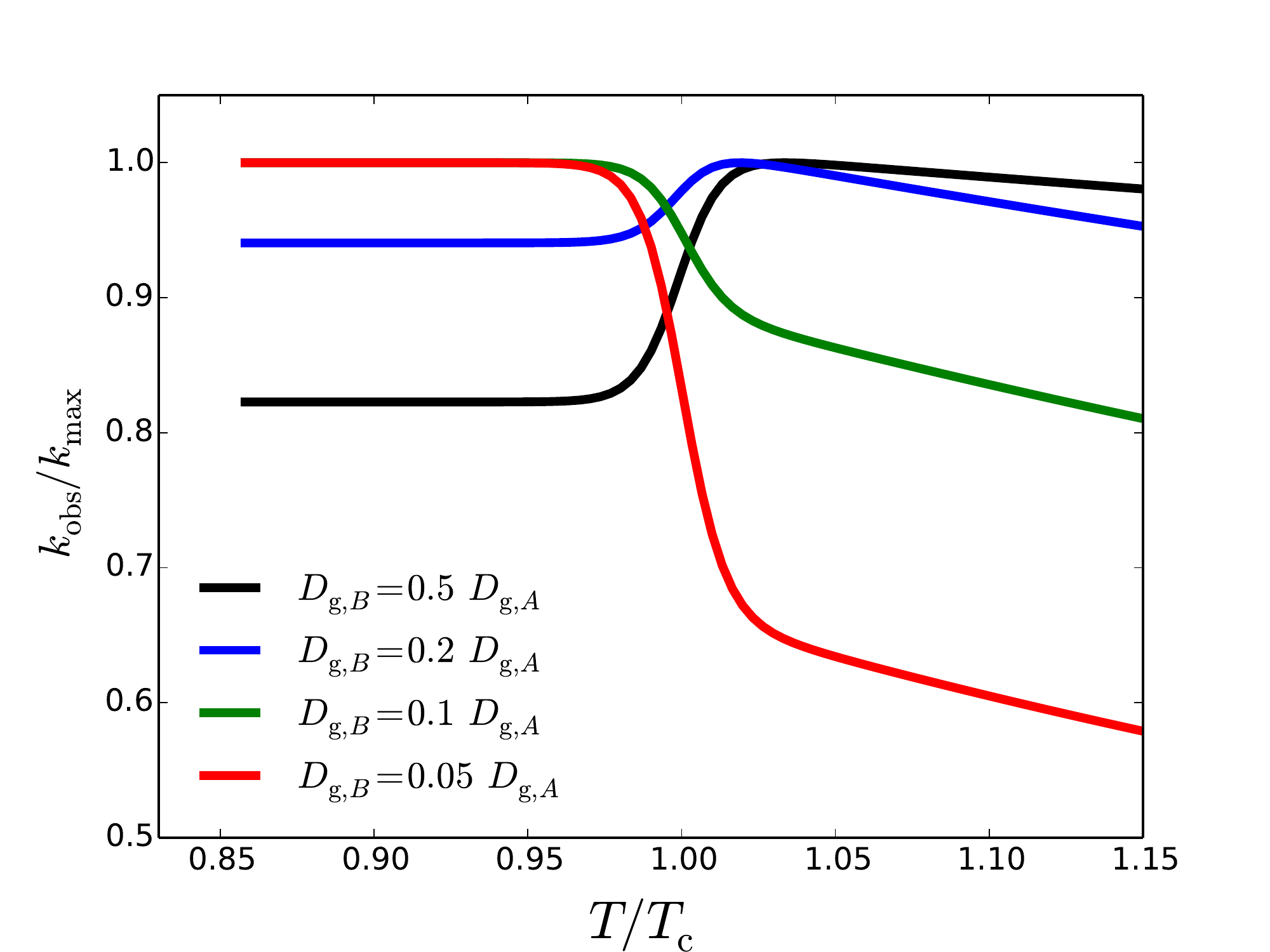}
\caption{Parameteric study of the influence of various factors on the reaction rate of nanoreactors as a function of temperature. A) For different $\Delta \DGsp = \DGsb - \DGsa$ (and fixed value for
\DGsb. B) At constant $\Delta \DGsp (<0)$, but changing the final value of \DGsb\,\,via its enthalpy $\Delta H_{{\rm sol},B}$. C) For varying $D_{B} / D_{A}$, i.e. for different drops in the diffusion coefficient in the collapsed state, in the case where $\Delta \DGsp$ is constant (and $<0$). This latter case shows how a large drop in diffusion coefficient can mask the effect of a decreasing free enthalpy. In this case, however, the apparent drop in $k_{\rm obs}$ will still be smaller than what would be calculated using an ideal diffusion model. For all but panel C), $D_{{\rm g},B} = 0.2\,D_{{\rm g},A} = D_{\rm bulk}$.}
\label{fig:parametric}
\end{center}
\end{figure}

\section{DISCUSSION \label{sec:discussion}}
In the following, we shortly present a parametric study to demonstrate the influence of the various parameters in our model. For ease of discussion, 
and to make contact with a known and well studied system, that is PNIPAM-based yolk-shell nanoreactors, we take the external parameter $\xi$ through which \DGt is tuned to be temperature.
Thus, the two two states of the gel are the low temperature swollen state $A$ and the high temperature collapsed state $B$, and the 
variation in temperature of \DGs is described via $\DGsc{,\alpha} = \Delta H_\alpha - T \Delta S_\alpha$ ($ \alpha=A,B $), with $\Delta H_\alpha, \Delta S_\alpha$ being constants.

In order to highlight interesting trends dependent on \DGsc, we take the case where the reaction rate is dominated by diffusion, 
i.e., $k_{\rm t}\approx k_{\rm D} << k_{\rm R}$. 
%
Fig.\,\ref{fig:parametric} shows $k_{\rm obs}$ vs temperature (normalized by the transition temperature) 
for three different scenarios. 
Panel A) shows $k_{\rm obs}$, normalized by its maximum value within the plotted temperature interval $k_{\rm max}$, 
for changing $\Delta\DGsp = \DGsa-\DGsb$ (i.e. the difference in solvation free enthalpy between the $A$ and $B$ state). 
When $\Delta \DGsp >0$, both terms appearing in $\scale$ contribute to a drop in the rate, but it should be noticed that this drop is much stronger than it would be observed based on the drop in the diffusion coefficient alone. Indeed, this fact can explain the large drop of the ``effective'' diffusion coefficient necessary 
to fit rate-vs-temperature curves in the model of Carregal-Romero \etal,\cite{marzan1}, which is not compatible with predictions from the hydrodynamic theories of diffusion that
should be valid in this experimental regime.\cite{masaro,amsden} 
When $\Delta \DGsp<0$, $k_{\rm D}$ will  depend on two contrasting effects: a decrease in the diffusion coefficient and an increase in the local 
concentration of reactant due to better solvation in the gel. In this case, for negative enough $\Delta\DGsc$, a jump to higher rates is observed close to the transition temperature.\\
Whereas $\Delta$\DGs dictates whether a jump or a drop in the rate occurs at the critical temperature, 
the absolute value of $k_{\rm obs}$ depends on the absolute value of \DGsp. 
In Panel B,  we fix $\Delta \DGsp (<0)$ by simply imposing $\Delta H_{\rm sol} = \Delta H_{{\rm sol},B} - \Delta H_{{\rm sol},A} = 5~{\rm {\rm kJ/mol}} $ and 
varying the absolute value of $\Delta H_{{\rm sol}, B}$ (note that we do not normalize $k_{\rm obs}$ here). An important fact emerges. Whenever the solvation free enthalpy is 
positive the drop of the rate is proportional to $\exp(|\Delta \DGsp|)$ (and is hence equal for all curves) because Eq.\,(\ref{eq:kdiff-nano}) is basically exact, and the rate is dominated by crossing of the polymer gel ($\tau_{\rm g} \gg \tau_\infty \gg \tau_{\rm np} $). However, when the solvation free enthalpy becomes negative $\tau_{\rm g}$ can decrease enough to be comparable or even below $\tau_\infty$. In this case, the rate is not proportional to $\exp(|\Delta \DGsp|)$ anymore and since $\tau_\infty$ has a weaker dependence on temperature than $\tau_{\rm g}$, $k_{\rm obs}$ becomes almost constant. This becomes evident for the curve where $\Delta H_{{\rm sol}, B} = -5~{\rm {\rm kJ/mol}}$, where the drop in the rate is just around $30\%$ compared to an almost 20-fold decrease for the other cases. 
Finally, panel C) refers to the case where we can  control the drop in the diffusion coefficient while keeping constant the solvation free enthalpy. 
In this case, we imposed the same $\Delta H_{{\rm sol},A(B)}$ and $\Delta S_{{\rm sol},A(B)}$ for all curves (giving $\Delta \Delta \bar{G_{\rm sol}}=-5~{\rm {\rm kJ/mol}}$), and we only  
change the diffusion coefficient in the $B$ state. A situation similar to that observed in panel A) is seen: The chosen change in \DGs would force 
an upward jump of the rate at the critical temperature, but this is counterbalanced by the drop expected due to 
a lower diffusion coefficient in the high-temperature, collapsed state. Depending on the relative magnitude of these two effects, 
different signatures are again observed.\\
One last thing to notice in Fig.\,\ref{fig:parametric} is the behavior of the rate as a function of temperature away from the transition region. 
Whenever this is dictated by $\tau_{\rm g}$ (i.e. when Eq.\,(\ref{eq:kdiff-nano}) is valid), the rate increases or decreases with temperature 
depending on the sign of the solvation enthalpy: a negative enthalpy means a decrease with temperature, whereas the opposite is observed for positive values, 
with the rate of decrease (i.e. the slope of the curve) being higher the higher the absolute value of $\Delta H_{\rm sol}$.\\ 
\begin{figure}[H]
\begin{center}
\includegraphics[width=0.45\textwidth]{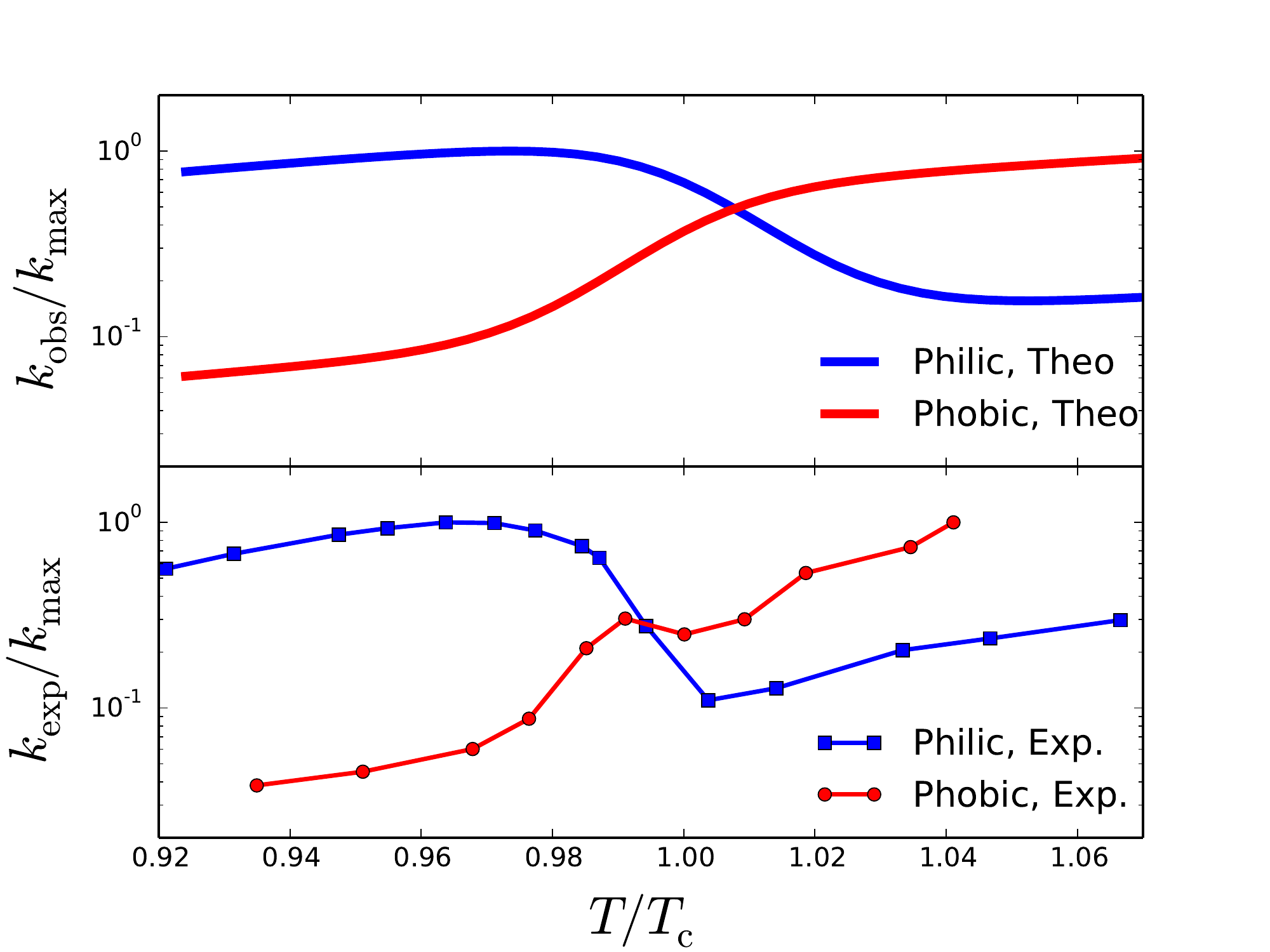}
\caption{Top: Normalized reaction rate as a function of temperature, in the case that $\scale$ decreases (red curve) or increases (blue curve) upon crossing \lcst. 
Note that since the diffusion coefficient in the gel is always expected to drop upon polymer collapse at \lcst, an increase in $\scale$ can be observed only if the solvation free enthalpy decreases, leading to a better interaction of the reacting molecule with the solvating gel environment, and hence to an increased concentration of reactants inside the gel. 
\DGs enters exponentially in the definition of $\scale$(see Eq.\,(\ref{eq:kdiff-nano})), hence even a decrease of a few $k_{\rm B} T$ can offset a large reduction in diffusion coefficient. 
On the contrary, an increase in \DGs can lead to an even larger reduction in the reaction rate than what would be expected by a simple reduction in the diffusion coefficient.}
\label{fig:discussion1}
\end{center}
\end{figure}
Evidence of the two qualitatively different behaviors we predict based on our theory can be found in recent experiments on different 
PNIPAM-based nanoreactor architectures,\cite{angewandte, marzan1} which we show in Fig.\,\ref{fig:discussion1}. In general, one expects that, 
for a hydrophilic molecule, the solvation free enthalpy is lower in the low-temperature, hydrophilic state of the gel compared to that in the high-temperature 
hydrophobic state. Hence, upon crossing \lcst from below there will be both a drop in the diffusion coefficient {\it and} an increase in the 
solvation free enthalpy that will exponentially reduce the reaction rate. For a  hydrophobic molecule the opposite behavior should be observed for the solvation free enthalpy, 
and an upward jump in the rate can be observed. 
These cases are in full agreement with the experimental data.\cite{marzan1, angewandte} In the first case, 
Carregal-Romero and coworkers looked at the reduction of hexacyanoferrate(III) ($\mathrm{Fe(CN)_6^{3-}}$), a strongly hydrophilic molecule due to its charge, 
by borohydride anions using a {\it core-shell} nanoreactor.\cite{marzan1} These workers indeed found a marked drop in the reaction rate
(blue squares in Fig.\,\ref{fig:discussion1}), much larger than what would be expected due to the drop in the diffusion coefficient as calculated using hydrodynamic 
theories that should well describe the relevant experimental regime.\cite{masaro,amsden}
Instead, Wu \etal used yolk-shell nanoreactors to control the reduction of 4-nitrobenzene using  borohydride anions ($\mathrm{BH_4^-}$). 
Going from below to above the \lcst, gel permeability is increased in this case due to a lower \DGs in the high-temperature state, inducing the observed jump in the reaction rate (red circles in Fig.\,\ref{fig:discussion1}, bottom) 
despite the decrease in the diffusion coefficient in the collapsed network.
An important aspect to point out is that the aforementioned reactions are red-ox reactions, and hence in principle of second order. 
However, in the experiments we compare to the concentration of $\mathrm{BH_4^-}$ was chosen in a way to make the reaction pseudo first order
(i.e. $\mathrm{BH_4^-}$  was present in large excess with respect to stoichiometric conditions), as also explicitly verified from the experimental kinetic data 
\cite{marzan1,angewandte,Wunder2010}. In this case, our theory is still fully applicable.\\

Before we conclude, let us now briefly discuss the main limitation of our model. Since we treat the reaction within the Debye picture, the polymer cage simply acts as a {\it fixed}, external field for the 
diffusing reactant. A more complete description instead would account for the fact that the state of the polymer, e.g. its local density, will adapt to the presence of the diffusing specie. Although description of the 
reaction kinetics including such a coupling is possible using techniques such as (classical) dynamic density functional theory \cite{ddft} or coarse-grained molecular dynamics simulations, 
the intrinsic complication arising in this more complex case would make the problem analytically intractable, and one would need to resort to numerical calculations. 
Our aim here was instead to provide a simple model to rationalize the likely origin of important trends observed in polymer-based nanoreactors, as we do. Certainly, a more quantitative description of the 
problem will require the use of such techniques, and we are already planning such simulations for the future.

\section{CONCLUSIONS}
We developed a model to describe the reaction rate observed in polymer-based, stimuli-responsive catalytic nanoreactors.
The theory combines a two-state thermodynamic model with the description of the reactants' diffusion which is based on 
Debye's theory of diffusion through an energy landscape. Model calculations highlight the importance of the solvation free enthalpy difference between the bulk solvent and the nanoreactor's polymeric 
cage. The theory predicts not only a sudden decrease in the observed rate, but also a possibility for rate enhancement, 
depending on the change in solvation free enthalpy at the swollen-to-collapse transition of the PNIPAM-based nanoreactors. 
Such rate enhancement has been observed in recent experiments\cite{angewandte,marzan1} and corroborates our description. 
The entire treatment demonstrates that nanoreactors can be used to enhance the selectivity of catalysis by nanoparticles. 

\section{ASSOCIATED CONTENT}
\subsection{Supplementary Information}
A step-by-step derivation for the full Debye's equation (including finite reaction rates), specifying all assumptions and limitations, is provided in the Supplementary Information.
This material is available free of charge via the Internet at http://pubs.acs.org.

\section{Acknowledgements}
S. A.-U. acknowledges the Alexander von Humboldt Foundation (AvH) for funding via an AvH Postdoctoral Research Fellowship.
Research in the J.D. group is supported by the Deutsche Forschungsgemeinschaft (DFG), the AvH, and the  ERC (European Research Council) Consolidator Grant with project number   646659 -- NANOREACTOR.

\providecommand*\mcitethebibliography{\thebibliography}
\csname @ifundefined\endcsname{endmcitethebibliography}
  {\let\endmcitethebibliography\endthebibliography}{}


\newpage

\section{Table of Content Graphics}

\begin{figure}[H]
\begin{center}
\includegraphics[width=0.95\textwidth]{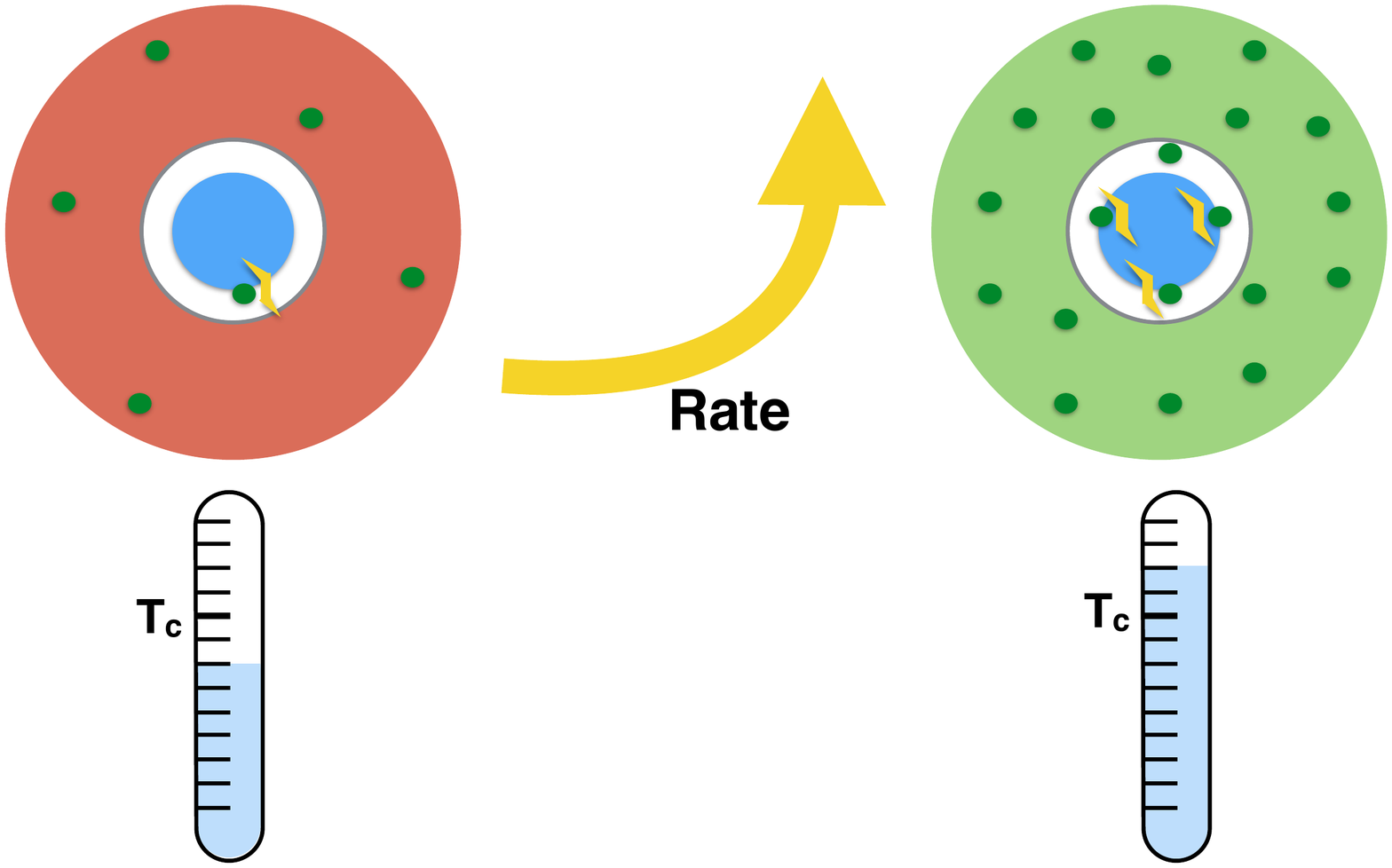}
\caption*{}
\end{center}
\end{figure}

\end{document}